\documentclass[aip,amsmath,amssymb,graphicx,floatfix,reprint]{revtex4-1}
\draft 
\usepackage{graphicx}
\usepackage{dcolumn}
\usepackage{bm}
\usepackage[usenames,dvipsnames]{color}
\usepackage[mathlines]{lineno}
\usepackage[utf8]{inputenc}
\usepackage[T1]{fontenc}
\usepackage{mathptmx}
\usepackage{etoolbox}
\usepackage[normalem]{ulem}
\begin{document}

\title{Influence of Temporal Variations in Plasma Conditions on the Electric Potential Near Self-Organized Dust Chains} 

\author{Katrina Vermillion}
\affiliation{CASPER, Baylor University, One Bear Place 97316, Waco, TX, 76798-7316, USA}
\email[Corresponding author: ]{katrina_vermillion1@baylor.edu}

\author{Dustin Sanford}
\affiliation{CASPER, Baylor University, One Bear Place 97316, Waco, TX, 76798-7316, USA}

\author{Lorin Matthews}
\affiliation{CASPER, Baylor University, One Bear Place 97316, Waco, TX, 76798-7316, USA}

\author{Peter Hartmann}
\affiliation{CASPER, Baylor University, One Bear Place 97316, Waco, TX, 76798-7316, USA}
\affiliation{Institute for Solid State Physics and Optics, Wigner Research Centre for Physics, PO Box 49, H-1525 Budapest, Hungary}

\author{Marlene Rosenberg}
\affiliation{Department of Electrical and Computer Engineering, University of California at San Diego, La Jolla, CA, 92093, USA}

\author{Evdokiya Kostadinova}
\affiliation{CASPER, Baylor University, One Bear Place 97316, Waco, TX, 76798-7316, USA}
\affiliation{Department of Physics, Leach Science Center, Auburn University, Auburn, AL, 36849, USA}

\author{Jorge Carmona-Reyes}
\affiliation{CASPER, Baylor University, One Bear Place 97316, Waco, TX, 76798-7316, USA}

\author{Truell Hyde}
\affiliation{CASPER, Baylor University, One Bear Place 97316, Waco, TX, 76798-7316, USA}

\author{Andrey M. Lipaev}
\affiliation{Joint Institute for High Temperatures, Russian Academy of Sciences, Izhorskaya 13/19, 125412 Moscow, Russia}

\author{Alexandr D. Usachev}
\affiliation{Joint Institute for High Temperatures, Russian Academy of Sciences, Izhorskaya 13/19, 125412 Moscow, Russia}

\author{Andrey V. Zobnin}
\affiliation{Joint Institute for High Temperatures, Russian Academy of Sciences, Izhorskaya 13/19, 125412 Moscow, Russia}

\author{Oleg F. Petrov}
\affiliation{Joint Institute for High Temperatures, Russian Academy of Sciences, Izhorskaya 13/19, 125412 Moscow, Russia}
\affiliation{Moscow Institute of Physics and Technology, Institutsky Lane 9, Dolgoprudny, Moscow Region 141700, Russia}

\author{Markus H. Thoma}
\affiliation{I. Physikalisches Institut, Justus-Liebig-Universit\"{a}t Gie{\ss}en, Heinrich-Buff-Ring 16, 35392 Gie{\ss}en, Germany}

\author{Mikhail Y. Pustylnik}
\affiliation{Institut f\"{u}r Materialphysik im Weltraum, Deutsches Zentrum f\"{u}r Luft- und Raumfahrt (DLR), Linder H\"{o}he, 51147 Cologne, Germany}

\author{Hubertus M. Thomas}
\affiliation{Institut f\"{u}r Materialphysik im Weltraum, Deutsches Zentrum f\"{u}r Luft- und Raumfahrt (DLR), Linder H\"{o}he, 51147 Cologne, Germany}

\author{Alexey Ovchinin}
\affiliation{Gagarin Research and Test Cosmonaut Training Center, 141160 Star City, Moscow Region, Russia}

\date{\today}

\begin{abstract}
\label{abstract}
The self-organization of dust grains into stable filamentary dust structures (or "chains") largely depends on dynamic interactions between the individual charged dust grains and the complex electric potential arising from the distribution of charges within the local plasma environment.  Recent studies have shown that the positive column of the gas discharge plasma in the Plasmakristall-4 (PK-4) experiment onboard the International Space Station (ISS) supports the presence of fast-moving ionization waves, which lead to variations of plasma parameters by up to an order of magnitude from the average background values.  The highly-variable environment resulting from ionization waves may have interesting implications for the dynamics and self-organization of dust particles, particularly concerning the formation and stability of dust chains.  Here we investigate the electric potential surrounding dust chains in the PK-4 by employing a molecular dynamics model of the dust and ions with boundary conditions supplied by a Particle-in-Cell with Monte Carlo collisions (PIC-MCC) simulation of the ionization waves.  The model is used to examine the effects of the plasma conditions within different regions of the ionization wave and compare the resulting dust structure to that obtained by employing the time-averaged plasma conditions.  Comparison between simulated dust chains and experimental data from the PK-4 shows that the  time-averaged plasma conditions do not accurately reproduce observed results for dust behavior, indicating that more careful treatment of plasma conditions in the presence of ionization waves is required.  It is further shown that commonly used analytic forms of the electric potential do not accurately describe the electric potential near charged dust grains under these plasma conditions.
\end{abstract}

\pacs{}
\maketitle 

\section{Introduction}
\label{intro}
The self-organization of dust grains in a flowing plasma offers insight into the interplay among charged species in a complex plasma environment.  In a complex plasma, micrometer-sized particles (also known as dust) acquire charge due to interactions with ions and electrons.  The resulting charge on a dust grain is typically negative in the absence of secondary thermionic or photoelectric emission due to the high mobility of electrons compared with the much more massive ions.  Ground-based experiments have studied the formation of dust structures in the sheath region of a plasma\cite{Kompaneets2007}, where a large vertical electric field is required to balance the gravitational force acting on the dust particles.  Ions drift relative to the more massive dust grains, with the ion drift velocity ($v_i$) corresponding to the strength of the applied electric field.  The trajectories of ions in the vicinity of a charged dust grain are deflected, leading to the build-up of ions downstream from the dust grain, forming a region of enhanced ion density called an ion wake.  The presence of positively charged ion wakes is thought to be responsible for the non-reciprocal interactions that have been observed between negatively charged dust grains in a flowing plasma\cite{Ivlev2015, Kliushnychenko2017}.

\begin{figure}
\includegraphics[width=\linewidth]{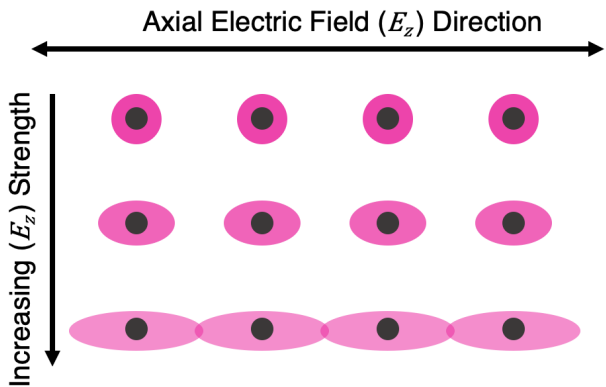}
\caption{Illustration showing the qualitative behavior of the shielding ion cloud surrounding dust grains as the alternating axial electric field strength ($E_z$) increases.  In the top row, representing $E_z=0$, the shielding cloud is spherical and close to the dust grain.  In the middle row, representing a moderate increase in $E_z$, the shielding ion clouds begin to stretch in the axial direction.  In the bottom row, representing $E_z$ above a critical value, the ion clouds are even more distorted and stretch across the entire region between individual dust grains.}
\label{fig:ioncloudcartoon}
\end{figure}

In contrast, microgravity dusty plasma experiments performed on parabolic flights \cite{Thoma2006} or on the International Space Station (currently through the Plasmakristall-4, PK-4, experiment)\cite{Fortov2005,Ivlev2011,Khrapak2016,Thoma2010,Pustylnik2016,thomas2019} enable investigations of dust dynamics in the near-absence of gravitational influence, where the smaller effects from the dust-dust, dust-ion, and drag forces are the primary drivers of the dust motion.  In microgravity experiments, the dust cloud can be levitated within the bulk of the discharge plasma where electric fields are weaker, and ion flow speeds are lower, than typically found in a plasma sheath.  In order to confine the dust within the field of view, the polarity of the DC discharge of the PK-4 device undergoes polarity switching at a frequency $\omega_{pi} >\omega_{ps} >\omega_{pd}$ (where $\omega_{pi}$ is the ion plasma frequency, $\omega_{ps}$ is the polarity switching frequency, and $\omega_{pd}$ is the dust plasma frequency). As a result, the regions of enhanced ion density are symmetrically stretched along the direction parallel to the external electric field (illustrated in Fig. \ref{fig:ioncloudcartoon}).  This allows for a more detailed examination of the interaction between dust grains and the surrounding plasma environment, especially as it relates to the formation of dust particle chains, dust density waves, and other interesting phenomena which have previously been reported \cite{Arp2012,Zobnin2018,Takahashi2019,Pustylnik2020,Dietz2021}.  The complex interactions between ion wakes and the charged dust grains are thought to guide the homogeneous-to-string transition\cite{Lipaev97,Fortov97,Melzer1996,Melzer1999}, resulting in behavior similar to that observed in conventional electrorheological (ER) fluids\cite{Ivlev2008}.

Recent ground-based observations of ionization waves under the conditions present in the PK-4 experiment have been reported\cite{Hartmann2020}, and the impact that these observed ionization waves can have upon the plasma parameters of interest for dust chain formation suggest that further investigation is warranted.  PIC-MCC simulations of the plasma at 40 Pa found that the ions thermalized with the neutral gas between ionization fronts, with a large increase in the ion drift velocity associated with the passing of an ionization wave\cite{Matthews2021}.  Previous work investigated the effect of the large axial electric field associated with the ionization waves on the stability of the dust chains using numerical simulations of the dust charging and dynamics in the thermalized plasma\cite{Matthews2021}.  The results were obtained by increasing the applied electric field, representing different possible averages of the electric field, while the temperature and density of electrons and ions were unchanged.  The increased electric field strengths imposed a drift velocity on top of the thermal motion of the ions obtained from the average plasma conditions between ionization wave peaks, while the use of constant ion and electron temperatures and plasma densities resulted in a constant shielding length for all electric field strengths considered.  It was found that the electric field strength corresponding to the ionization wave front was more favorable to dust chain formation than a homogeneous plasma described by the average plasma conditions. 

In this work, the behavior of dust and ions in the PK-4 experiment at 60 Pa is investigated in order to match the conditions used in later experimental campaigns.  PIC-MCC simulations of the plasma at 60 Pa found that the thermal velocity of ions in the plasma remained about 12$\%$ higher than expected if the ions were to thermalize with the neutral gas, and electron and ion temperatures were found to vary continuously\cite{Hartmann2020}.  As a result, the current investigation was designed to incorporate the variations in electron and ion temperatures and plasma densities, which allows the effect of changing screening length on the resulting dust chains to be probed.  The simulation of the ion and dust dynamics also allows examination of the electric potential near interacting charged dust grains in the case of an anisotropic plasma with ion-neutral collisions.  The configuration of the electric potential gives insight into the stability of the dust structures formed in the dynamic plasma context.

This paper is organized as follows.  An overview of the PK-4 Experiment and the PIC-MCC model of the PK-4 discharge are given in Section \ref{background}.  A discussion of the plasma conditions selected for examination in the MD model of the dust and ions are given in Section \ref{methods}.  The MD model for dust and ions is described in Section \ref{model}, and includes details regarding the treatment of ions, dust dynamics, and dust grain charging.  Comparison of the resulting dust structure, electric potential, and ion wake effects at different time periods within ionization waves are presented in Section \ref{results}, with a discussion of the results in Section \ref{discussion} and concluding remarks in Section \ref{conclusion}.

\begin{figure} [b]
\includegraphics[width=\linewidth]{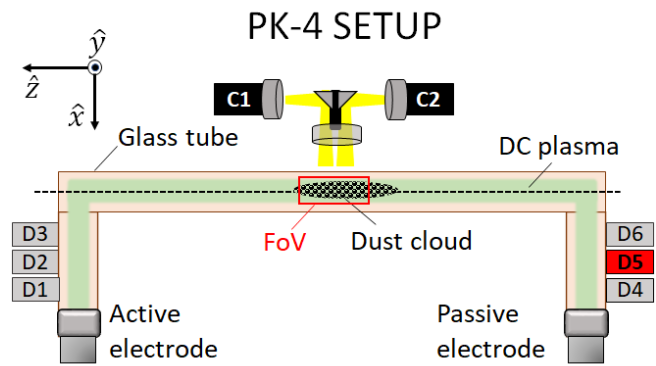}
\caption{Diagram of the PK-4 ISS experimental setup illustrating the $\pi$-shaped discharge tube and relative locations of the particle observation cameras (C1 and C2), field of view (FoV), dust dispensers (D1-D6), and the active and passive electrodes.  Dispenser 5 (D5), shown highlighted in red, was used in the experiment modeled here to inject dust particles of diameter 3.38 $\mu m$.}
\label{fig:pk4_setup}
\end{figure}

\section{Background}
\label{background}
The Plasmakristall experiments have allowed the study of three-dimensional complex plasmas in microgravity since their first introduction on board the Mir space station in the late 1990's\cite{MorfillBook2013}.  The most recent iteration, the PK-4, has implemented a long axial DC discharge tube (instead of the compact radio-frequency, RF, plasma used in previous iterations of the Plasmakristall experiments) to further aid investigations of the liquid state of complex plasmas\cite{MorfillBook2013}.  This section will introduce the PK-4 experiment on the ISS  and the PIC-MCC model of the PK-4 DC plasma\cite{Hartmann2020}.

\subsection{PK-4 ISS Experiment}
\label{PK4}
The PK-4 experiment on the International Space Station consists of a 30-mm-diameter $\pi$-shaped cylindrical glass tube, high voltage power supply, RF coil which can be moved along the tube axis, dust dispensers, and both manipulation and illumination lasers.  The central portion of the main tube is used for imaging by a plasma glow observation camera as well as particle observation cameras.  A diagram of the PK-4 experimental setup is included for reference in Fig. \ref{fig:pk4_setup}.  In the particular experiment analyzed in this work, the dust is trapped within the field of view of the observation cameras using polarity switching of the DC electric field.  A more detailed description of the PK-4 experiment and capabilities can be found in [\onlinecite{Pustylnik2016}].

\subsection{PIC Model of PK-4}
\label{PIC model}
Previous theoretical results of dust dynamics have been obtained using the assumption that the positive column of the discharge plasma is homogeneous along the axial direction on timescales relevant to the dust dynamics\cite{Sukhinin2013}.  Although DC plasma discharges are known to contain self-excited oscillations over a large range of frequencies and varied plasma conditions\cite{Rohlena1972}, the frame rates used to record the PK-4 ISS glow limit observations to those processes which occur on ms or longer timescales\cite{Pustylnik2016}.  Video data collected from the PK-4 experiment appears to be homogeneous, though the presence of fast-moving ionization waves cannot be excluded.  

The 2D particle-in-cell with Monte Carlo collisions (PIC-MCC) model\cite{Donko2021} of local plasma parameters in the PK-4 system used in [\onlinecite{Hartmann2020}] revealed the presence of fast-moving ionization waves with a phase velocity in the range of $500-1200$ m/s.  The PIC-MCC model uses the simplified geometry of a 400-mm-long straight discharge tube with an inner diameter of 30 mm, and the cathode and anode placed at each end.  

The PIC-MCC simulation was validated against the ground-based experiment BUD-DC, which has a straight cylindrical discharge closely matching the simplified geometry used in the PIC-MCC model while keeping the system as close as possible to the PK-4 in terms of discharge tube diameter, gas type, and range of pressures\cite{Hartmann2020}.  Experimental data collected from the BUD-DC with an exposure time of $40$ ms shows that global discharge structure and current dependence are properly captured by the PIC-MCC simulation.  However, the PIC-MCC simulation also revealed the presence of fast-moving ionization waves.  This finding was confirmed by discharge light emission sequences collected using a high-speed CCD camera with frame rates of $50, 000$ fps in a second ground-based experiment, the PK-4 BU.  The PK-4 BU is a ground-based replica of the PK-4 consisting of a 30-mm-diameter $\pi$-shaped cylindrical quartz-glass tube, a power supply capable of operation in polarity switching mode with switching frequencies up to 1 000 Hz, and a movable RF coil\cite{Schmidt2020}. 

\begin{figure}[b]
\includegraphics[width=\linewidth]{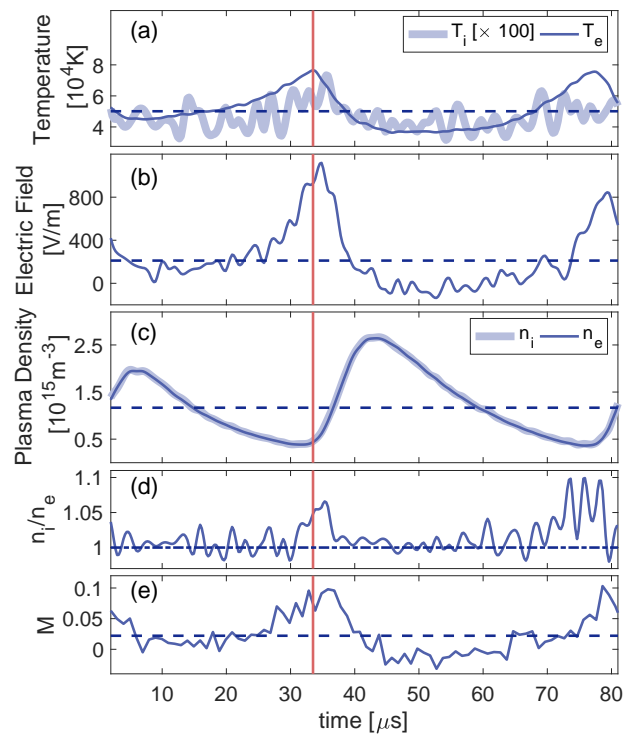}
\caption{Results from the PIC-MCC model illustrating time-varying (a) electron and ion temperatures, $T_e$ and $T_i$, (b) axial electric field, $E_z$, (c) electron and ion number densities, $n_e$ and $n_i$, (d) the ratio of $n_i$ to $n_e$, and (e) the Mach number, $M$.  The associated time-averaged value of the parameters are indicated by a dashed line corresponding to the values $\langle T_e\rangle  = 50107$ K = 4.32 eV, $ \langle E_z \rangle  = 211.1 $ V/m, $\langle n_e\rangle  = 1.18 \times 10^{15}$ m$^{-3}$, and $\langle M \rangle = 0.022$, respectively.  The dot-dashed line in (d) corresponds to $n_i/n_e = 1$.  The vertical red line in each panel marks the time of the first peak in temperature.  Discharge parameters are: neon gas, pressure = 60 Pa, current = 2 mA.}
\label{fig:PICdata}
\end{figure}

\subsection{Results from PIC Model of PK-4}
\label{PIC results}
The main feature revealed by the PIC-MCC model is that conditions within the positive column of the PK-4 discharge lead to large, fast-moving ionization waves occurring on timescales on the order of 10 kHz. The local plasma parameters within the ionization waves vary by as much as an order of magnitude from the time-averaged values, as shown in Fig. \ref{fig:PICdata}.  Further, the frequency of the ionization waves is high enough to prevent the ions from thermalizing with the neutral gas between subsequent peaks of the ionization waves, which results in continuously varying plasma conditions.  This is in contrast with the PIC-MCC results obtained at 40 Pa used for the simulation in [\onlinecite{Matthews2021}], where the regions between ionization waves were shown to be relatively homogeneous and punctuated only by brief variations in plasma conditions.

The heating of electrons and ions (Fig. \ref{fig:PICdata}a) is accompanied by large variations in the axial electric field (Fig. \ref{fig:PICdata}b), followed by a depletion of both ions and electrons as the ionization wave passes (Fig. \ref{fig:PICdata}c).  The time delay between peaks in the different plasma parameters is logical, considering that the charge separation provided by the imbalance of electrons and ions produces the large electric field observed in the ionization wave.  This offset between maximum and minimum values in various parameters leads to interesting consequences for the formation of ion wakes and dust structures, and will be discussed below.  

\subsection{Implications for String Formation}
\label{string formation}
Previous investigations of dust particle chain formation have focused on time-averaged plasma parameters due to the dust particle’s large inertia and, therefore, large response time compared with other plasma species\cite{Arp2012} (with dust typically moving on a millisecond timescale while ions respond on a microsecond timescale).  However, as illustrated in Fig. \ref{fig:PICdata}, ionization waves in the plasma lead to large oscillations in the axial electric field, plasma number densities, and the electron and ion temperatures (based on the mean kinetic energies) on a microsecond timescale\cite{Hartmann2020}, which in combination impact the formation of ion wakes in complex plasmas.  Since the character of the ion wake changes dramatically, further investigation is needed to determine if this is translated into the dust motion.

Interactions between dust grains and a flowing plasma have been characterized using the Mach number, $M = v_{i}/c_{s}$, where $v_{i}$ is the ion drift velocity and $c_s = \sqrt{k_B T_e/m_i}$ is the ion sound speed (as in [\onlinecite{Hutchinson2012}]). The Mach number (representing the ion flow speed) is shown in Fig. \ref{fig:PICdata}e.  Previous investigation of intergrain forces in low-Mach plasma wakes using a particle-in-cell code\cite{Hutchinson2012} found that wake enhancements were not present when the flow velocity is less than $0.3 c_s$, given $T_e/T_i = 100$ (as is the case in the current investigation, shown in Fig. \ref{fig:PICdata}a).  Moreover, it was noted that grain-aligning forces would be negligible and likely would not be responsible for the formation of dust chains at low Mach numbers.  As shown in Fig. \ref{fig:PICdata}e, the range of ion flow speeds found are all much less than this threshold value suggesting that, in this case, the dust should be unlikely to form strings.  However, results from PK-4 experiments show that the modeled conditions do lead to the formation of long (tens of particles) stable dust strings, suggesting that the established criteria for string formation may be incomplete.

\section{Methods}
\label{methods}
\subsection{Data Selection}
\label{data selection}
The current investigation seeks to isolate the influence that the changing plasma conditions have on the stability of dust chains. In this study we model a small volume within the dust cloud, focusing on the dynamics of a single chain of 20 dust grains with a radial confinement force designed to mimic the effect of surrounding dust chains. The structure of the electric potential surrounding the dust chain, including the contributions from the charged dust grains and the localized concentrations of ions, is examined at different times within the ionization wave.  The electron and ion temperatures, which influence dust grain charging and the ion dynamics, reach values within ionization waves that are two to six times the background values (see Fig. \ref{fig:PICdata} a and c).  The plasma densities also impact the formation of wake structures around the charged dust grains, which in turn influences both the formation and the stability of the dust chain.  The axial electric field (Fig. \ref{fig:PICdata}b) which drives the ion flow leading to wake formation and dust chain stability, reaches peak values in the ionization waves as large as 20 times the value between peaks.  Evidently, each of these parameters impact stability and formation of dust chains, and the specific contribution from any individual component could be difficult to isolate.  However, the time offset between peak values seen in Fig. \ref{fig:PICdata} results in the plasma parameters varying in time with respect to each other, which allows for closer observation of the impact each parameter may have.

\begin{figure}[t]
\includegraphics[width=\linewidth]{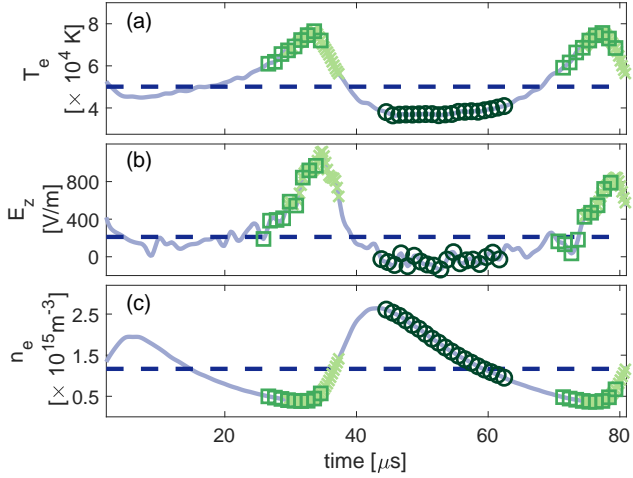}
\caption{Results of the PIC-MCC model showing (as a solid line) the time-varying (a) electron temperature, $T_e$, (b) axial electric field, $E_z$, and (c) electron number density, $n_e$.  The data points selected for the various averaging schemes are denoted by: \emph{Between $E_z$ Peaks} (dark green circles), \emph{Rising $E_z$ and Minimum $n_i$} (medium green squares), and \emph{FWHM $E_z$ Peaks} (light green x's).  For comparison, the \emph{Full Time Average} value of each variable is shown as a dashed horizontal line extending over the full extent in each plot.}
\label{fig:inputdatacomp}
\end{figure}

Specific regions from the time history are selected based on the behavior of the electric field for the present investigation.  The plasma conditions within the identified regions of interest are used in the MD simulation to investigate their effect on string formation, and results are subsequently compared to dust dynamics where the full time-averaged plasma conditions are used.  The different plasma conditions are: Case (1) \emph{Full Time Average}, where the values of each parameter averaged over a time span (440 $\mu$s) covering several ionization waves are considered; Case (2) \emph{Between $E_z$ Peaks}, where only data between the large peaks in the axial electric field are considered, representing a region of minimum ion flow; Case (3) \emph{Rising $E_z$ and Minimum $n_i$}, where data within the rise of the axial electric field are considered, corresponding to a time when electron and ion number densities are depleted; Case (4) \emph{Full-Width Half-Maximum $E_z$ Peaks}, where the data is averaged over the FWHM of peaks in the axial electric field, representing the maximum ion flow.  A representative time interval indicating the data selected for each of the averaging cases is illustrated in Fig. \ref{fig:inputdatacomp}, and the resulting averaged plasma parameters for each case are listed in Table \ref{tab:params}.

\begin{table*}
    \caption{Plasma parameters used for each of the four averaging cases.  Note that $+(-)$ values for $E_z$ and $v_{z,ion}$ indicate parallel (anti-parallel) orientation relative to the tube axis.}
    \label{tab:params}
    \begin{ruledtabular}
    \begin{tabular}{lccccc}

       & 1. Full Time Average & 2. Between $E_z$ Peaks & 3. Rising $E_z$ and Min. $n_i$ & 4. FWHM $E_z$ Peaks \\
       ${E_z}$ [V/m] & 211.1 & -45.2 & 474.3 & 814.8 \\
       ${n_{e0}}$ [m$^{-3}$] & 1.17e15 & 1.58e15 & 4.32e14 & 6.29e14 \\
       ${n_{i0}}$ [m$^{-3}$] & 1.18e15 & 1.59e15 & 4.43e14 & 6.48e14 \\
       ${T_e}$ [K] & 50107 & 37549 & 65330 & 69767 \\
       ${T_i}$ [K] & 466 & 425 & 509 & 565 \\
       ${v_{z,ion}}$ [m/s] & 111.5 & -29.8 & 231.7 & 377.3 \\
       Mach [M] & 0.022 & 0.008 & 0.043 & 0.070 \\
       $\omega_{pi}$ [MHz] & 10.1 & 11.7 & 6.2 & 7.5 \\
      
    \end{tabular}
\end{ruledtabular}
\end{table*}

\section{Simulating Dust and Ion Motion}
\label{model}
The numerical model used to simulate the dust and ion dynamics is DRIAD (Dynamic Response of Ions And Dust).  This model is described in detail in [\onlinecite{Matthews2020}], and here we provide only a brief overview of the components relevant to the current investigation.  DRIAD is a molecular dynamics simulation which resolves dust and ion motion each at their own relevant timescales using a timestep of $\Delta t_d = 10^{-4} $ seconds for the dust and $\Delta t_i = 10^{-9} $ seconds for the ions.     

The motion of ions and dust grains are governed by their respective equations of motion.  For an ion of mass $m_i$, the equation of motion is given by 
\begin{equation} m_i \ddot{\vec{r}} = \vec{F}_{ij} + \vec{F}_{iD} + \vec{F}_E(z) + \vec{F}_{bound}(r,z) + \vec{F}_{in}. \label{eq:ionEqMotion}\end{equation}
The electrostatic force between ions $i$ and $j$, $\vec{F}_{ij}$, is derived from a Yukawa potential, where the electrons provide shielding, resulting in the expression \begin{equation} \vec{F}_{ij} = \sum_{i \neq j} \frac{q_i q_j}{4\pi \epsilon_0 r_{ij}^3}\left( 1 + \frac{r_{ij}}{\lambda_{De}}\right) \mathrm{exp}\left(-\frac{r_{ij}}{\lambda_{De}}\right)\vec{r}_{ij},\label{eq:forceii}\end{equation}
where $q_i$ ($q_j$) denotes the charge of particle $i$ ($j$), $r_{ij}$ denotes the distance between particles $i$ and $j$, $\lambda_{De} = \sqrt{\epsilon_0 k_B T_e/(n_e e^2)}$ is the electron Debye length with $T_e$, $n_e$, and $e$ representing the temperature, number density, and charge of electrons, respectively.  The force on the ions due to dust grains, $\vec{F}_{iD}$, is derived from a Coulomb potential (as the electrons are depleted in the vicinity of the negatively charged dust\cite{Piel2017}) \begin{equation} \vec{F}_{iD} = \sum_d \frac{q_i Q_d}{4 \pi \epsilon_0 r_{id}^3}\vec{r}_{id}, \label{eq:forceid}\end{equation} 
where the sum is taken over the $d$ dust grains, $Q_d$ represents the charge on the dust grain, and $\vec{r}_{id}$ is the distance between an ion ($i$) and dust grain ($d$).  The treatment of ion-dust and ion-ion forces follows the Molecular Asymmetric Dynamics (MAD) code developed by Piel\cite{Piel2017}, previously shown to yield reasonable results for potential distributions and interparticle forces when compared to PIC simulations.  In the asymmetric treatment of dust and ions, the force ($\vec{F}_{iD}$) on the ions by the charged dust are calculated with the bare Coulomb potential while the force on the dust by the ions ($\vec{F}_{di}$) is calculated using a shielded Yukawa interaction to bridge the limiting cases of motion close to a dust grain where $|e\Phi | \gg k_B T_e$ (and electrons are depleted), and sufficiently far from the dust grain that $|e\Phi | \ll k_B T_e$ and ions are shielded by electrons.

The force on the ions due to the external axial electric field, $\vec{E}(z)$, which switches directions at the polarity switching frequency of 500 Hz, is $\vec{F}_E(z) = q_i \vec{E}(z)$.  The ions outside the simulation region exert a force on the ions within the simulation through $\vec{F}_{bound}(r,z) = q_i \vec{E}_{bound}(r,z)$.  The electric field due to ions outside the simulation region, $\vec{E}_{bound}(r,z)$, is calculated as the negative gradient of the potential in a cylindrical cavity, found by numerical calculation of the potential of uniformly distributed ions within a cylinder, which is then subtracted from a constant potential.
The ion-neutral collisions ($\vec{F}_{in}$) are treated using the null-collision method \cite{Skullerud1968,Donko2011}, with the isotropic and backscattering cross section collision data taken from the Phelps database (hosted by LXCat project)\cite{Carbone2021}.

The equation of motion for a dust grain with mass $m_d$ is given by 
\begin{equation} m_d \ddot{\vec{r}} = \vec{F}_{di} + \vec{F}_{dD} + \vec{F}_E(z) + \vec{F}_{drag} + \vec{F}_{B} + \vec{F}_{C}.\label{eq:dusteqmotion}\end{equation}
The force of ions on the dust grains, $\vec{F}_{di}$, is calculated from a Yukawa interaction, similar to Eq. \ref{eq:forceii}.  At each dust timestep, the cumulative force of ions acting on a dust grain is averaged over the elapsed ion timesteps.  The force between dust grains, $\vec{F}_{dD}$, is a Coulomb interaction, similar to Eq. \ref{eq:forceid}, as the screening by ions is included in the expression for $\vec{F}_{di}$.  The axial electric field provides a force $\vec{F}_E(z)=Q_d \vec{E}(z)$. The drag force, $\vec{F}_{drag}$, is the neutral gas drag.  The dust is immersed in a thermal bath which is implemented using the force $\vec{F}_B$.

The confinement force, $\vec{F}_C$, simulates the effect of neighboring dust chains outside the simulation region.  The confinement is included using the expression \begin{equation} \vec{F}_C = \omega_d^2 Q_d\vec{r} ,\label{eq:forceconfine}\end{equation} where the confining strength, $\omega_d^2$ is proportional to  \begin{equation}\omega_d^2 \propto \frac{C\tilde{Q}}{4 \pi \epsilon_0 \triangle^2}\left(1+\frac{\triangle}{\lambda_{De}}\right)\mathrm{exp}\left(-\frac{\triangle}{\lambda_{De}}\right),\label{eq:omega}\end{equation} with $\tilde{Q}$ and $\triangle$ representing the expected average dust charge and inter-grain separation estimated from results in the PK-4 experiment.  The constant $C$ is used to scale the confining force provided by surrounding chains, and is based on the expected inter-chain spacing and average dust grain charge.  To aid in comparison, the value of $\omega_d^2$ was set to the same value for each of the four cases considered (3.96$\times10^{6}$ V/m$^2$), and the confinement force of Eq. \ref{eq:forceconfine} is ultimately proportional to the charge on a dust grain in the simulation.

In the experiment, the confinement force ($\vec{F}_C$) would be provided by discrete dust grains present in the neighboring dust chains.  Although the influence from individual dust grains does lead to a "ripple" in the resulting potential along the axial direction, simulation of the Yukawa potential resulting from regularly spaced neighboring dust chains revealed maximum deviations from a smooth potential to be on the order of a few hundredths of a mV.  The use of a smooth confinement force is therefore considered to be a reasonable approximation for the current investigation.

Note that no axial confinement force is imposed by the simulation other than the polarity switching of the axial electric field.  The dust grains are allowed to spread in the z-direction to reach their equilibrium axial interparticle spacing.

The dust grain charging in DRIAD is in response to electron and ion currents; as the electrons are not modeled directly, they are assumed to be Boltzmann-distributed, and the electron current, $I_e$, is calculated using orbital motion limited (OML) theory: \begin{equation} I_e = 4\pi a^2 n_e e\left(\frac{k_B T_e}{2\pi m_e}\right)^{1/2}exp\left(\frac{e\phi_d}{k_B T_e}\right),\label{eq:electroncurrent}\end{equation} where $n_e$, $m_e$, and $T_e$ are the electron density, mass, and temperature, respectively, $k_B$ is the Boltzmann constant, and $\phi_d = Q_d/(4\pi \epsilon_0 a)$ is the surface potential of the dust grain.  The ion current is comprised of ions which cross the ion collection radius, $b_c$, \begin{equation} b_c = a\left(1-\frac{2q_i\phi_d}{m_i V_s^2}\right)^{1/2}, \label{eq:ioncollradius}\end{equation}  $V_s$ is the characteristic velocity of the ions \begin{equation} V_s = \left( \frac{8k_B T_i}{\pi m_i}+ v_i^2\right)^{1/2},\label{eq:ionvelocity}\end{equation} and $v_i$ is the drift speed of an ion, which may conveniently be reported as a fraction of the sound speed $v_i = M c_s$.  Calculating the electron current using the OML theory and the ion current by tracking the number of ions collected by each dust grain results in a charge that is different from the OML theory in that (1) the charge is affected by the ion flow, (2) the ion current to the dust grain is enhanced as a result of ion-neutral collisions, and (3) the inhomogeneity in the ion distribution is naturally taken into account.  The results of the DRIAD simulation are found to be consistent with predictions based on OML theory which have been corrected to account for flowing ions and the presence of ion-neutral collisions\cite{Matthews2021}.

\section{Results}
\label{results}
Dust charging and dynamics for 20 dust grains were simulated using the plasma conditions shown in Table \ref{tab:params} for each of the four cases.  All of the simulations used neon gas with neutral gas pressure $p = 60$ Pa, neutral gas temperature $T= 295$ K, electric field polarity switching frequency of 500 Hz, dust radius $a = 1.69$ $ \mu$m, and dust mass density corresponding to that of melamine formaldehyde (1.51 g/cm$^3$).  The simulation region is a cylinder of radius 840 $\mu$m and length 8050 $\mu$m, which provides enough space for the dust grains to reach the expected average interparticle spacing determined from the PK-4 experiment with at least one electron Debye length between the axial ends of the simulation and the outermost dust grain.  Dust grains are prevented from leaving the simulation region by a strong confinement force at the simulation boundaries.  The dust was initially placed in a region roughly 450 $\mu$m in diameter near the center of the simulation region.  The dust cloud expanded, and was allowed to evolve for $1.2$ seconds, with the dust in each case reaching equilibrium after approximately $1.0$ seconds.  The dust configuration in each case was found to be stable during the final 200 ms (which covers at least 2 dust plasma periods ($\tau_d$), where $\tau_d = 2\pi \sqrt{\epsilon_0 m_d/(n_{d}Q_d^2)}$ and dust number density $n_d \approx 10^{10}$ m$^{-3}$), and the final 100 ms of data was used for the results presented below.  The same plasma conditions used in Cases 1 - 4 of Table \ref{tab:params} were also used to model a single stationary dust grain, where only dust grain charging and the motions of ions are considered.  The ion density and electric potential near the isolated dust grains are compared with those from the simulation results for the chains formed by the dynamic dust grains, and relevant results are provided in the following sections.  The final average grain charges and interparticle separations are listed in Table \ref{tab:lengths}.  The magnitude of the grain charge increases with the electron temperatures and the ion flow speed, which also leads to a greater interparticle separation due to the stronger repulsion between grains.  However, there is not a linear relationship between the grain charge and interparticle spacing, due to the distribution of ion density and the resulting electric potential, as explained below.

\subsection{Ion Density}
\label{ionden}
\begin{figure}[t]
\includegraphics[width=\linewidth]{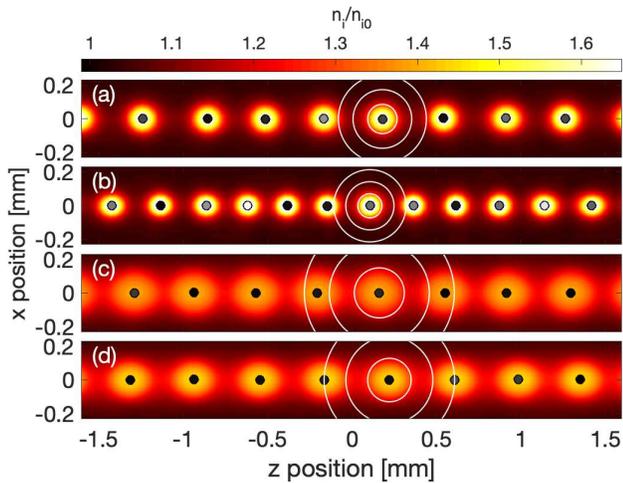}
\caption{Plot of ion density for (a) Case 1: \emph{Full Time Average}, $n_{i0} = 1.18\times 10^{15}$ m$^{-3}$, (b) Case 2: \emph{Between $E_z$ Peaks}, $n_{i0} = 1.58\times 10^{15}$ m$^{-3}$, (c) Case 3: \emph{Rising $E_z$ and Minimum $n_i$}, $n_{i0} = 4.32\times 10^{14}$ m$^{-3}$, (d) Case 4: \emph{FWHM $E_z$ Peaks}, $n_{i0} = 6.29\times 10^{14}$ m$^{-3}$.  The colorbar corresponds to the ion density normalized to the background ion density far from the dust: $n_i / n_{i0}$.  The shade of the dust grain indicates distance along the y axis (in or out of the page), with black indicating dust grains positioned in the xz-plane and white indicating dust grains 0.25 $\lambda_{De}$ from the xz-plane.  The three concentric white circles in each plot indicate the distances 2, 4, and 6 $\lambda_{Di}$ from the location of one of the dust grains.}
\label{fig:ionden}
\end{figure}

The ion densities averaged over the final 0.1 seconds for each of the four cases simulating 20 moving dust grains are shown in Fig. \ref{fig:ionden}.  Note that the plot is zoomed in on the central part of each chain so that the details in the radial direction are visible.  The dust grains are shown as grayscale dots, where the color of the dot indicates the magnitude of the distance from the xz-plane (out-of-plane direction), with black indicating dust grains positioned in the xz-plane and white indicating a displacement of 0.25 $\lambda_{De}$, the maximum out-of-plane distance found at equilibrium.  The colorbar indicates the ion density normalized to the background ion density far from the dust, $n_i/n_{i0}$, where values for $n_{i0}$ for each case are given in Table \ref{tab:params}.  From the data shown in Fig. \ref{fig:ionden}, it is clear that the different cases lead to different ion wake structures, as expected since the extent and location of an ion wake formed in the presence of a charged dust grain depends on the external electric field and resultant ion drift velocities.  

\begin{table*}
    \caption{Size of the cylindrical simulation region expressed in Debye lengths, final average interparticle spacing, and final average dust grain charge for each case.}
    \label{tab:lengths}
    \begin{ruledtabular}
    \begin{tabular}{lccccc}

     & & 1. Full Time Average & 2. Between $E_z$ Peaks  & 3. Rising $E_z$ and Min. $n_i$ & 4. FWHM $E_z$ Peaks \\
    \hline
    Electron Debye Length, $\lambda_{De}$ &[$\mu$m] & 450.3 & 335.5 & 838.5 & 715.8\\
    Ion Debye Length, $\lambda_{Di}$ &[$\mu$m] & 43.4 & 35.7 & 74.0 & 64.4 \\
    \\
    Radius $(838.5 \mu$m$)$ &$[\lambda_{De}]$ & 1.9 & 2.5 & 1.0 & 1.2\\
    Length $(8050 \mu$m$)$ &$[\lambda_{De}]$  & 17.9 & 23.8 & 9.6 & 11.3\\
    \\
    Interparticle Spacing, d &[$\mu$m] & 345 & 246 & 363 & 370 \\
     &[$\lambda_{De}$] & 0.77 & 0.73 & 0.43 & 0.52 \\
     &[$\lambda_{Di}$] & 7.9 & 6.9 & 4.9 & 5.7 \\
    \\
    Charge &[e$^-$] & 2260 & 2040 & 2270 & 2630 \\

    \end{tabular}
\end{ruledtabular}
\end{table*}

In Case 1 (which uses the full time average of plasma parameters, Fig. \ref{fig:ionden}a) and in Case 2 (which is averaged over the minimal external electric field between ionization wavefronts, Fig. \ref{fig:ionden}b), the dust forms a somewhat aligned chain with most of the disorder seen in the out-of-plane direction, and the large ion density maxima centered around the dust grains.  In Case 3 (averaged over the rise in external electric field, Fig. \ref{fig:ionden}c), the ion density maxima are less intense and are elongated between the dust grains, and the chain structure shows considerably less displacement out-of-plane.  Case 4 (averaged over the FWHM peaks of the external electric field), Fig. \ref{fig:ionden}d, shows results similar to that of Case 3, with even greater enhancement of ion density between grains.  The results shown in Fig. \ref{fig:ionden} lead to two broad classifications of the four averaging cases: low ion flow with distinct ion clouds collected around each dust grain (as in Cases 1 and 2), and high ion flow resulting in elongation and merging of the ion clouds (as in Cases 3 and 4).

The various ion and electron temperatures and plasma densities result in a different ion Debye length for each of the four averaging cases examined (the Debye lengths for each case are provided in Table \ref{tab:lengths}).  Using the white circles (indicating radii of 2, 4, and 6$\lambda_{Di}$) in Fig. \ref{fig:ionden} as a guide, it is evident that a significant transition has occurred between Cases 1 and 2 and Cases 3 and 4.  In Cases 3 and 4, the average interparticle spacing is smaller ($\sim$5-6 $\lambda_{Di}$) than in Cases 1 and 2 ($\sim$7-8 $\lambda_{Di}$).  The close proximity of dust grains with respect to the ion Debye length in Cases 3 and 4 are paired with a smearing-out of the ion density enhancement along the axial direction, while the more isolated dust grains in Cases 1 and 2 correspond to highly localized regions of ion density enhancement.

\subsection{Electric Potential}
\label{electricpotential}

\begin{figure}[b]
\includegraphics[width=\linewidth]{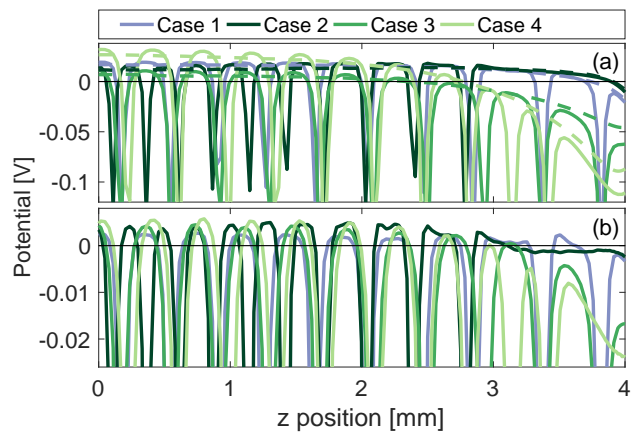}
\caption{Plots showing (a) the total electric potential from the DRIAD simulation along the z-axis (solid lines) and the fit to the axial potential taken at a radial distance of $9\lambda_{Di}$ from the z-axis (dashed lines), and (b) the resulting total electric potential after subtracting the fit to the axial potential.  The horizontal black line indicates the level $V=0$ in both plots.}
\label{fig:roll_calcs}
\end{figure}

 The simulation results for the electric potential have a large variation in the background potential along the z-axis, illustrated in Fig. \ref{fig:roll_calcs}a, as the ions accumulate near the dust chain and normalized ion density is near unity at the ends of the chains. This overall variation in the background potential is expected to be an artifact of the simulation, and is eliminated by subtracting a fit to the potential along a line parallel to the z-axis at a radial distance of $9\lambda_{Di}$.  At this distance the axial potential profile is essentially unaffected by the presence of the dust grains (as defined by the normalized ion density being equal to one).  The total potential of the ions and dust (indicated by a solid line) and the fit to the background potential (indicated by a dot-dashed line) are shown in Fig. \ref{fig:roll_calcs}a.  The resulting potentials along the z-axis are shown in Fig. \ref{fig:roll_calcs}b, and the more uniform potential along the z-axis allows for a direct comparison between the four averaging cases considered.

\begin{figure}[t]
\includegraphics[width=\linewidth]{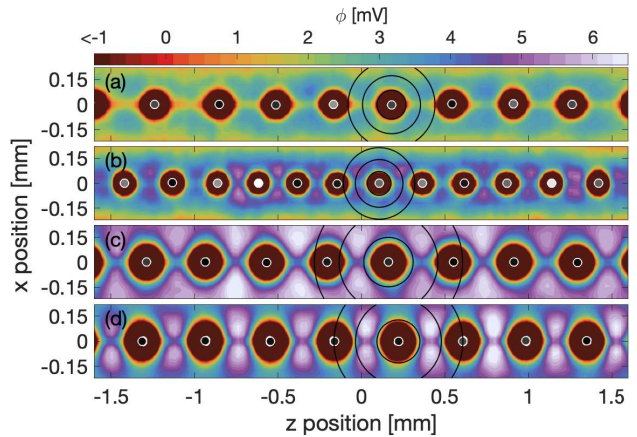}
\caption{Plots showing electric potential for (a) Case 1: \emph{Full Time Average}, (b) Case 2: \emph{Between $E_z$ Peaks}, (c) Case 3: \emph{Rising $E_z$ and Minimum $n_i$}, (d) Case 4: \emph{FWHM $E_z$ Peaks} averaging cases. The shade of the dust grain indicates distance from the y-axis (out of the page), with black indicating dust grains positioned on the xz-plane and white indicating dust grains farther from the xz-plane.  The three concentric black circles in each plot indicate the distances 2, 4, and 6 $\lambda_{Di}$ from the location of one of the dust grains.}
\label{fig:totalpotential}
\end{figure}
 
 The electric potential data, a combination of the negatively charged dust and positive ion wakes, $\phi = \phi_{ion} + \phi_{dust}$, averaged over the final $0.1$ seconds are shown in Fig. \ref{fig:totalpotential} for each of the four cases with 20 moving dust grains.  As in Fig. \ref{fig:ionden}, the view is zoomed in on the central portion of the chain to show the detail in the radial direction, and the grayscale color of the dust grains indicates the distance from the xz-plane (out of the page).  The relative scale of ion Debye lengths for each case are indicated by the concentric black circles surrounding the central dust grain, with radii corresponding to 2, 4, and 6$\lambda_{Di}$.  Each case shows a negative potential well at the dust grain locations, with the extent and depth of the potential well correlated to the magnitude of the charge on the dust grain, which increases with ion flow speed.  The maximum positive potential is located between 3-5 $\lambda_{Di}$ from a dust grain, and moves closer to a grain as the spacing of the dust grains (in ion Debye lengths) decreases. The magnitude of the positive potential between dust grains in Case 3 (Fig. \ref{fig:totalpotential}c) is $\approx 75\%$ larger than Case 1 (Fig. \ref{fig:totalpotential}a) and $\approx 30\%$ larger than in Case 2 (Fig. \ref{fig:totalpotential}b).  Case 4 (Fig. \ref{fig:totalpotential}d), which corresponds to the largest dust grain charge, has positive potential values $\approx 90\%$ and $\approx 40\%$ larger than Cases 1 and 2, respectively.

The large interparticle spacing in Case 1 (Fig. \ref{fig:totalpotential}a) relative to the ion Debye length results in minimal overlap of the potential from individual dust grains, allowing the form of the overall potential structure to closely resemble that of an isolated dust grain. In contrast, the close proximity of dust grains relative to the ion Debye lengths in Case 3 and Case 4 (Fig. \ref{fig:totalpotential}c and d, respectively) lead to interesting behavior that differs from a simple superposition of the electric potential from each individual dust grain, which will be discussed further in Section \ref{discussion}.

\begin{figure}[t]
\includegraphics[width=\linewidth]{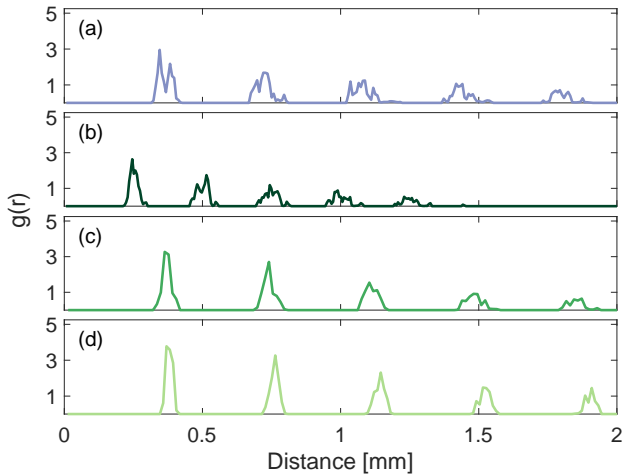}
\caption{Plot of g(r) pair correlation functions for (a) Case 1: \emph{Full Time Average}, (b) Case 2: \emph{Between $E_z$ Peaks}, (c) Case 3: \emph{Rising $E_z$ and Minimum $n_i$}, (d) Case 4: \emph{FWHM $E_z$ Peaks} averaging cases.}
\label{fig:gr_function}
\end{figure}

\subsection{Chain Structure}
\label{grtalk}
The final average interparticle spacing in each case are indicated in Table \ref{tab:lengths} measured in $\mu m$, $\lambda_{De}$, and $\lambda_{Di}$.  The final interparticle spacing found in Case 1 (representing the time-averaged plasma conditions) was 345 $\mu m$, which is larger than the 230 $\mu m$ average interparticle spacing observed in the experiment.  Cases 3 and 4 also have larger interparticle spacing (363 and 370 $\mu m$, respectively) than that observed in the experiment.  Case 2 most closely matches the experiment with d = 246 $\mu m$.  It is evident that using the time-averaged plasma conditions results in an over-prediction of the particle charge and a correspondingly large interparticle spacing.

To characterize the order seen in the dust chains in each case, the pair correlation function, $g(r)= (\langle\sum_{i\neq 0}\delta (\vec{r}-\vec{r}_i)\rangle/\rho$, where $\rho = $ (number of particles)$/$(volume), was calculated for each averaging case, shown in Fig. \ref{fig:gr_function}.  The pair correlation functions are calculated for the central 12 dust grains using the dust positions over the final 0.1 seconds of simulation time. Order is indicated by well-defined, sharp peaks in the $g(r)$ function at distances corresponding to integer multiples of the average interparticle spacing.  As shown in Fig. \ref{fig:gr_function}, the greatest correlation is seen for Cases 3 and 4 (which correspond to smaller interparticle separation measured in Debye lengths).  Cases 1 and 2 show the least order in the $g(r)$ function, which correlates with the increased disorder in the out-of-plane direction noted in these cases and larger interparticle separation measured in Debye lengths.

\begin{figure}
\includegraphics[width=\linewidth]{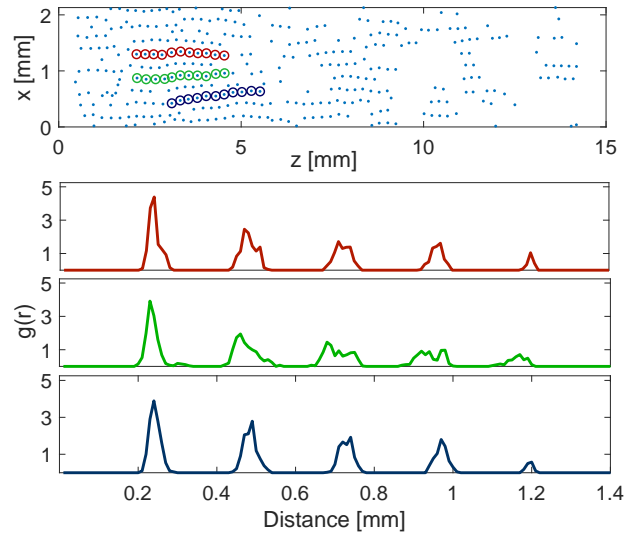}
\caption{Results from analysis of video data obtained during Campaign 7 of the PK-4 ISS experiment showing (top panel) chains chosen for analysis and (bottom 3 panels) the corresponding pair correlation functions for the three selected chains, averaged over 30 frames.  Colors used to highlight chain particles in the top panel correspond to the color of the $g(r)$ function in the bottom three panels.}
\label{fig:pk4_comp}
\end{figure}

The pair correlation functions also allow for comparison between the DRIAD simulation results and results obtained from microgravity experiments.  Parameters used for the current simulation closely match experimental conditions during Campaign 7 of the PK-4 experiment, which were 70 Pa neon gas, 0.7 mA current, 500 Hz electric field switching frequency, and 1.69 $\mu$m radius dust grains.  Video data were collected at 71.4 fps, from which 2D images of the dust cloud were analyzed using particle tracking over 90 frames.  Three chains comprised of 11 dust grains each were selected from the dust cloud, and are highlighted in the top pane of Fig. \ref{fig:pk4_comp}.  These particular chains were selected for analysis because they were similar in length to those analyzed in the simulation and remained intact over the entire time interval.  The pair correlation functions for the three selected chains were calculated and averaged over 30 frames (equivalent to $\approx$ 420 ms), and results are shown in the bottom three panes of Fig. \ref{fig:pk4_comp}.  The color of the pair correlation plots in the bottom three panes of Fig. \ref{fig:pk4_comp} correspond to the color used to highlight the dust grains in the image of the dust cloud shown in the top pane.

Qualitatively comparing the results in Fig. \ref{fig:pk4_comp} with those in Fig. \ref{fig:gr_function}, the shape of the first five nearest neighbor peaks in the pair correlation function obtained from experimental data most closely match the results in Case 3 (which is averaged over the rise in external electric field) and Case 4 (which is averaged over the FWHM peaks of the external electric field).  The plasma conditions in Cases 3 and 4 lead to the greatest order in the dust chains.  Although the average interparticle spacing seen in Case 2 is closer to the PK-4 experimental results, the chain formed in Case 2 shows little order compared with the long range order seen in the experimental results.

\section{Discussion}
\label{discussion}
The structure of the ion wake around dust grains has been shown to impact the formation and stability of dust chains in a complex plasma\cite{Ivlev2008,Matthews2020}.  From the g(r) plots it is apparent that Cases 3 and 4 have undergone a transition to a more ordered structure than that found in the results from Cases 1 and 2.  The increase in order is correlated with the transition from spherical, isolated ion density clouds surrounding the dust in Cases 1 and 2 to the ion density enhancement becoming more uniform along the length of the dust chain (parallel to the electric field), and decreased interparticle spacing relative to the ion Debye length.

The increased interparticle spacing in the simulated results (Fig. \ref{fig:gr_function}) when compared with the experimental results (Fig. \ref{fig:pk4_comp}) is likely due to increased dust charge due to lower dust number density ($n_d$) in the DRIAD simulation.  The visualized portion of the experimental dust cloud used for analysis in Fig. \ref{fig:pk4_comp} contains 347 dust particles in a 2.5 mm $\times$ 15 mm window, and is illuminated by a laser sheet with a FWHM extent that varies between 40-220 $\mu$m (characterized in [\onlinecite{Pustylnik2016}]).  This yields a dust number density within the PK-4 ISS experiment on the order of $n_d \approx 10^{10}-10^{11}$ m$^{-3}$, while Langmuir probe measurements in the dust-free discharge\cite{Pustylnik2016} show an electron density on the order of $n_e\approx 4\times 10^{14}$ m$^{-3}$.  The charge on dust grains in microgravity conditions at 60 Pa are theoretically estimated in [\onlinecite{Antonova2019}] to be $Q_d\approx 2\times 10^3$ e$^-$, indicating that as much as 50$\%$ of the electrons present in the dust cloud reside on the dust grain surface.  It is therefore logical to conclude that the neighboring dust chains present in the experiment lead to a decrease in the electron density within the dust cloud (dependent on the dust number density) that is unaccounted for in the current simulation.  If this depletion of the electrons were taken into account, the charge on the dust grains in the simulation conditions would be reduced.  However, even with this limitation, the qualitative behavior of the pair correlation function seen in the experiment most closely matches the simulation results in Cases 3 and 4. 

Models describing the electric potential surrounding charged particles in a plasma are useful for both planning experiments and analyzing their results, but the assumptions used in the derivation of a given model and its range of applicability must be carefully considered.  The large variation of ion density, electric potential, and the resulting order characterized by $g(r)$ in Cases 2-4 from the full time average plasma conditions in Case 1 suggest that electric potential models which are based on averaged plasma conditions may not fully describe the environment in the PK-4 experiment with the presence of the recently observed ionization waves.

\subsection{Comparison with Common Electric Potential Equations}
\label{potcomps}
The simplest model for the electric potential surrounding a point charge is the \emph{Coulomb potential} (Eq. \ref{eq:forceid}).  This form of potential is only dependent on the distance $r$ from the point charge, and therefore fails to capture any effects resulting from local non-neutrality of the charge distribution close to a dust grain.  To account for the changes in charge distribution near a dust grain, it is necessary to use the slightly more complex form of electric potential developed by H. Yukawa\cite{Yukawa1934}, which is typically adapted for use in the field of complex plasmas as the \emph{Yukawa potential} (Eq. \ref{eq:forceii}).  The Yukawa potential (also sometimes referred to as a Debye-H\"{u}ckel potential) is dependent not only on the distance from the point charge, but also on the temperature and number density of the charged species within the plasma through the term $\lambda_D$ in the exponent.  

While the Yukawa potential does improve upon the Coulomb potential by taking into account the shielding provided by mobile charges in the plasma, it fails to account for the changes in potential structure that arise from a flowing plasma.  A potential model based upon the Bhatnagar-Gross-Krook collision operator has been developed by R. Kompaneets\cite{KompaneetsDiss}, which has the form \begin{equation}\phi(\vec{r}) = \frac{Q}{|\vec{r}|^3}F_{BGK}(\theta) + o\left(\frac{1}{|\vec{r}|^3}\right),\label{eq:kompPot}\end{equation} with 
\begin{eqnarray}F_{BGK}(\theta) =&& -\sqrt{\frac{8}{\pi}}\frac{u\lambda_{D,T_n}^2}{v_{T_n}}\mathrm{cos}\theta + \left(2-\frac{\pi}{2}\right)\frac{u^2\lambda_{D,T_n}^2}{v_{T_n}^2}\nonumber \\
& & \times\left(1-3\mathrm{cos}^2\theta\right) +o(u^2),\label{eq:kompPot2}\end{eqnarray} where $\lambda_{D,T_n}=\sqrt{T_n/(4\pi n_0e^2)}$, $u$ represents the drift velocity, $v_{T_n} = \sqrt{T_n/m_n}$, and $\theta$ represents the angle between $\vec{r}$ and the direction of ion drift.  This form has been adapted to account for the symmetric ion wake present in a plasma subjected to an alternating external electric field (as in [\onlinecite{Ivlev2010}]), yielding the form referred to here as the \emph{Multipole Expansion Potential}:
\begin{equation} V(r,\theta) = Q\left[\frac{\mathrm{e}^{-r/\lambda_{Di}}}{r} - 0.43\frac{M_T^2\lambda_{Di}^2}{r^3}\left(3\mathrm{cos}^2\theta - 1\right)\right],\label{eq:ivlevpot}\end{equation}
where $M_T=u_i/v_{Ti}$ is defined as the "thermal" Mach number, and $\theta$ represents the angle between the external electric field driving the ion drift and $\vec{r}$.

The analytic forms of the electric potential were developed by considering either a single, isolated probe or the interaction between two probes separated by large distances (relative to the ion Debye length).  The development of the Multipole Expansion potential relies on the underlying assumption that the thermal Mach number is small ($M_T \ll 1$) and that the distance from the dust grain is much larger than the ion Debye length ($r \gg \lambda_{Di}$), as well as the assumptions inherent in the BGK collision operator, namely that the collision frequency is velocity independent and only charge exchange collisions are considered.

\begin{figure}
\includegraphics[width=\linewidth]{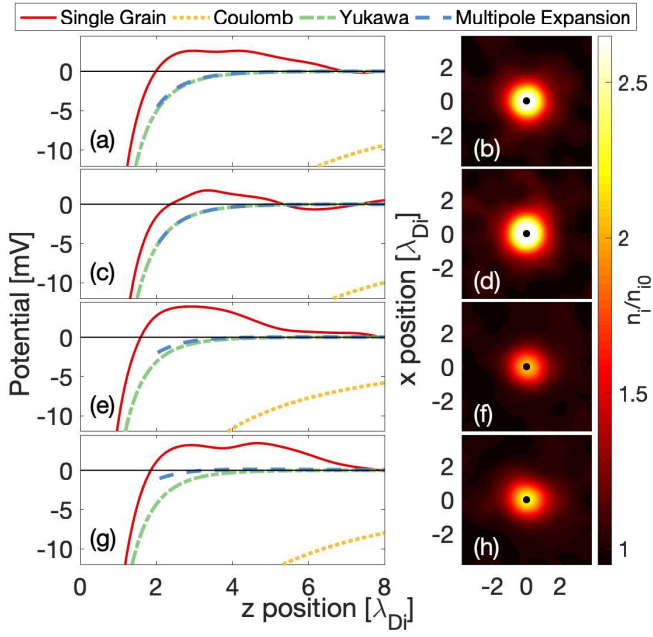}
\caption{Results from simulations of a single dust grain in plasma conditions corresponding to Case 1: \emph{Full Time Average} (a and b), Case 2: \emph{Between $E_z$ Peaks} (c and d), Case 3: \emph{Rising $E_z$ and Minimum $n_i$} (e and f), and Case 4: \emph{FWHM $E_z$ Peaks} (g and h).  The total electric potential (left column) from simulations (solid red line) is compared with the Coulomb Potential (dotted yellow line), Yukawa Potential (dot-dashed green line), and Multipole Expansion potential (dashed blue line).  The ion density (right column) values are normalized to the background ion density ($n_{i0}$) for each case, which are (b) Case 1: $n_{i0} = 1.18\times 10^{15}$ m$^{-3}$, (d) Case 2: $n_{i0} = 1.58\times 10^{15}$ m$^{-3}$, (f) Case 3: $n_{i0} = 4.32\times 10^{14}$ m$^{-3}$, and (h) Case 4: $n_{i0} = 6.29\times 10^{14}$ m$^{-3}$.}
\label{fig:analytic_pot_comp}
\end{figure}

To compare directly with the analytic potential forms, the plasma conditions used in Cases 1 - 4 were applied to a single, stationary dust grain located at the origin.  The plasma conditions in the four cases are identical to those from Table \ref{tab:params}, discussed in Sections \ref{methods} and \ref{model}, with the only change being that the dust grain does not move.  The ion density data from the single, stationary dust grain simulations are shown in the right column of Fig. \ref{fig:analytic_pot_comp}.  The magnitude of ion density enhancement surrounding the dust grains is largest for Case 2 (Fig. \ref{fig:analytic_pot_comp}d), as is the case for the simulations containing 20 dust grains.  However, asymmetry in the ion density enhancement in Case 3 (Fig. \ref{fig:analytic_pot_comp}f) and Case 4 (Fig. \ref{fig:analytic_pot_comp}h) is much less pronounced than in the 20 dust grain simulations (Fig. \ref{fig:ionden}a and b, respectively).  This result suggests that it is the interaction between closely spaced neighboring dust grains (relative to the scale of $\lambda_{Di}$) in addition to the plasma conditions which lead to the formation of the elongated ion wake structures found in Cases 3 and 4 (Fig. \ref{fig:ionden}c and d, respectively), and not the result of the plasma conditions alone.  Simulation results for the electric potential of a single stationary dust grain along the axial direction are shown in the left column of Fig. \ref{fig:analytic_pot_comp}, and compared to the analytic potential forms. 

\begin{figure}
\includegraphics[width=\linewidth]{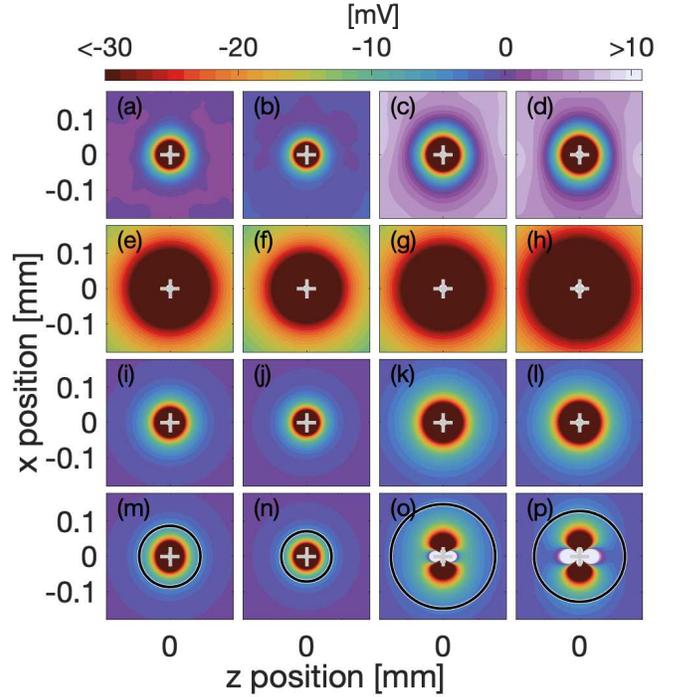}
\caption{Comparison of the simulation results for a single, isolated dust grain (top row: a, b, c, and d) with the \emph{Coulomb Potential} (second row: e, f, g, and h), \emph{Yukawa Potential} (third row: i, j, k, and l), and the \emph{Multipole Expansion Potential} (bottom row: m, n, o, and p).  The conditions used for the single grain simulations correspond to the conditions in Table \ref{tab:params} for Case 1 (first column), Case 2 (second column), Case 3 (third column), and Case 4 (fourth column).  The black circles in the bottom row (m-p) correspond to the radial distance of 2$\lambda_{Di}$ from the dust grain, which is indicated by the gray cross ('+') in each plot.}
\label{fig:single_grain_grid}
\end{figure}

In all four cases it is seen that the Yukawa potential and Multipole Expansion potential agree more closely with the simulation results and with each other than does the Coulomb potential.  However, both the Yukawa and Multipole Expansion forms predict a more negative potential than that obtained from the simulation results within a distance $r<\sim 5\lambda_{Di}$.  Fig. \ref{fig:analytic_pot_comp} shows that the difference between the Coulomb potential and the other potentials decrease as $\lambda_{Di}$ increases, while the difference between the Yukawa potential and Multipole Expansion potential decreases with decreasing Mach number.  To account for the breakdown in the validity of the Multipole Expansion form of the potential as $r\rightarrow 0$, the region $r < 2\lambda_{Di}$ has been omitted from the line plots in Fig. \ref{fig:analytic_pot_comp}.  The Coulomb potential predicts a larger negative potential than the simulation results due to the lack of an exponential shielding term.

\begin{figure}
\includegraphics[width=\linewidth]{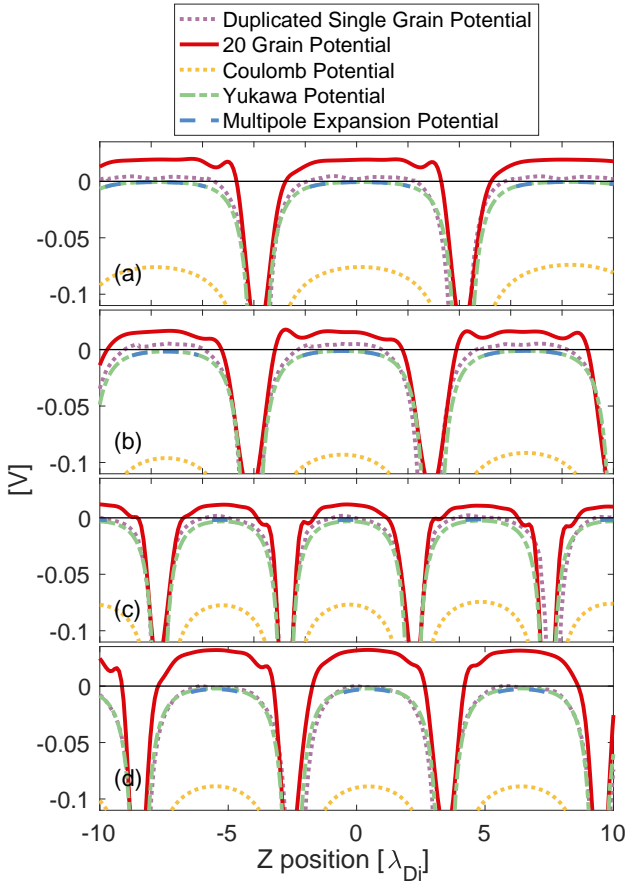}
\caption{Comparison of the superimposed single dust grain simulation results (dotted pink line) with the 20 dust grain simulation results (solid red line), \emph{Coulomb Potential} (dotted yellow line), \emph{Yukawa Potential} (dot-dashed green line), and the \emph{Multipole Expansion Potential} (dashed blue line).  Results are shown for (a) Case 1, \emph{Full Time Average}, (b) Case 2, \emph{Between $E_z$ Peaks}, (c) Case 3, \emph{Rising $E_z$ and Minimum $n_i$}, and (d) Case 4, \emph{FWHM $E_z$ Peaks}.}
\label{fig:analytic_pot_comp_full_chain}
\end{figure}

\begin{figure*}[t]
\includegraphics[width=\linewidth]{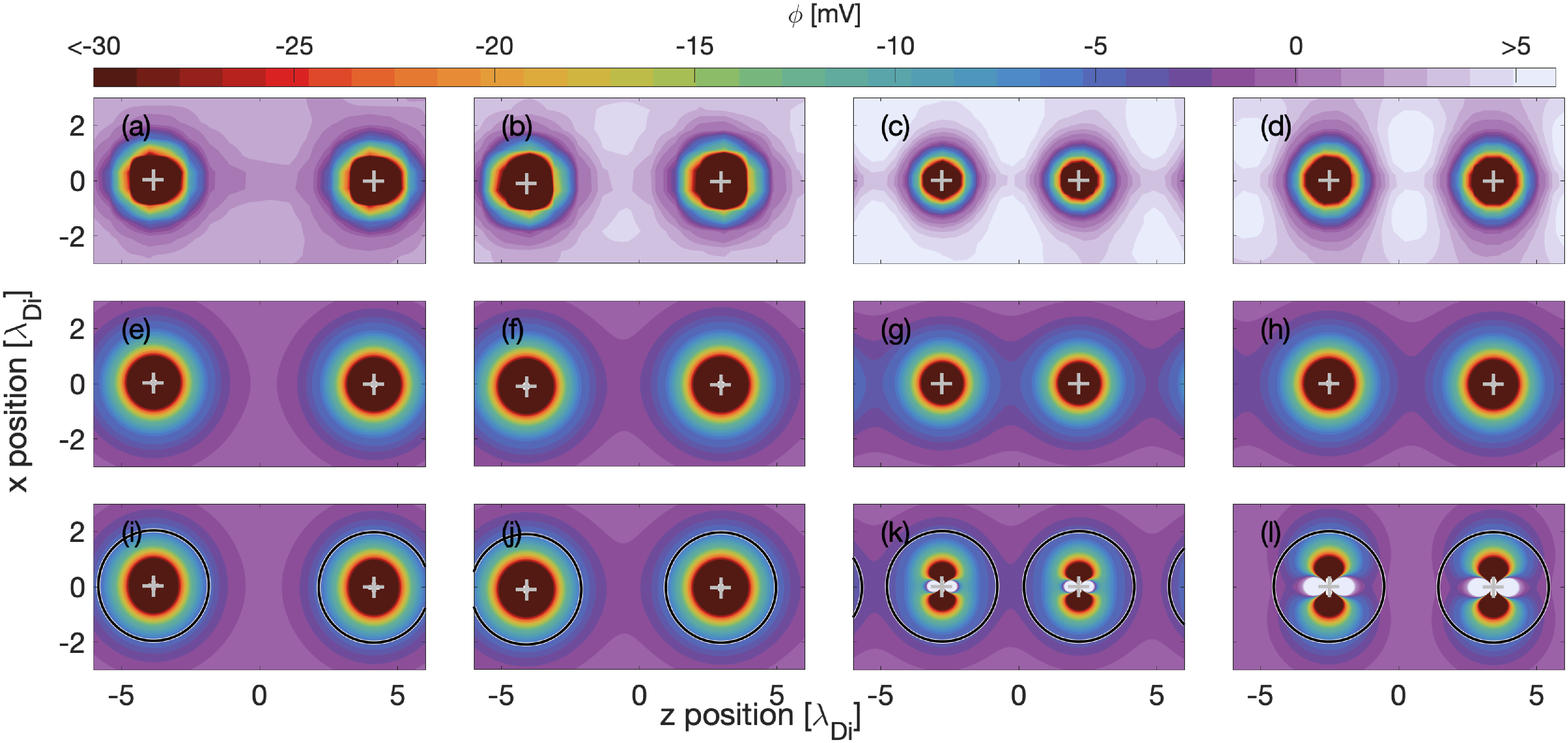}
\caption{Comparison of electric potential from the simulation (top row, a-d) with the analytic potential forms discussed in Section VI: the \emph{Yukawa Potential} is shown in the middle row (e-h), and the \emph{Multipole Expansion Potential} is shown in the bottom row (i-l).  The first column on the left (a, e, and i) shows Case 1 (\emph{Full Time Average}), the second column (b,f, and j) shows Case 2 (\emph{Between $E_z$ Peaks}), the third column (c, g, and k) shows Case 3 (\emph{Rising $E_z$ and Minimum $n_i$}), and the fourth column (d, h, and l) shows Case 4 (\emph{FWHM $E_z$ Peaks}).  Positions of dust grains are marked with a gray cross (+).  Black circles in i-l indicate the radius 2$\lambda_{Di}$ from dust grains.}
\label{fig:all_pot_comp}
\end{figure*}

Contour plots showing comparisons between the potential forms and simulation results from each of the four averaging cases with single, stationary dust grains are shown in Fig. \ref{fig:single_grain_grid}.  The region corresponding to 2$\lambda_{Di}$ from the dust grains is indicated on plots Fig. \ref{fig:single_grain_grid}m-p by the black circles, which correspond to the minimum radial extent considered valid for the Multipole Expansion potential in Eq. \ref{eq:ivlevpot}.  The asymmetry in the Multipole Expansion potential for Cases 3 and 4 (thought to contribute to the transition to flow-aligned dust chains\cite{Ivlev2011}) is evident only within the region $r< 2\lambda_{Di}$, which is excluded by the requirement that distance be much larger than the ion Debye length.

One may ask how applicable these results are to the case where more than one dust grain is present, and whether the case of multiple dust grains may be adequately represented by a superposition of the potentials of many single dust grains.  To address this question, the results of the simulations with 20 moving dust grains are compared with the Coulomb potential, Yukawa potential, Multipole Expansion potential, and with the single dust grain simulation results superimposed at each of the 20 dust grain locations, shown in Fig. \ref{fig:analytic_pot_comp_full_chain}.  The Multipole Expansion Potential is shown only in the regions farther than 2$\lambda_{Di}$ from any dust grains, to accommodate the restricted range of validity of this potential form.  As expected, the Coulomb potential form shows the greatest deviation from the simulation results for a 20 dust grain chain, with a more negative potential between dust grains for all of the plasma conditions examined.  The agreement between the superposition of single dust grain potentials (dotted pink line) and the potential from the 20 dust grain simulation (solid red line) is best for Case 2 (with the lowest axial electric field, ion drift velocity, and dust grain charge).  The agreement worsens as the axial electric field, ion drift velocity, and dust grain charge increase, a finding which is consistent with previous studies of the interaction between dust grains with anisotropic ion wakes\cite{Ivlev2008,Yaroshenko2021}.

Contour plots of the electric potential comparing the results from the simulations of 20 dust grains with the Yukawa and Multipole Potential forms are provided in Fig. \ref{fig:all_pot_comp}. The simulation results in Fig. \ref{fig:all_pot_comp}a-d correspond to the results shown in Fig. \ref{fig:totalpotential}, zoomed in on a smaller axial range to allow for closer comparison with the analytic potential forms.  The Yukawa potentials corresponding to each of the four averaging cases are shown in Fig. \ref{fig:all_pot_comp}e-h, and contour plots of the Multipole Expansion potential are shown in Fig. \ref{fig:all_pot_comp}i-l.  The region corresponding to 2$\lambda_{Di}$ from the dust grains is indicated on plots Fig. \ref{fig:all_pot_comp}i-l by the black circles, which correspond to the minimum radial extent considered valid for the Multipole Expansion potential in Eq. \ref{eq:ivlevpot}.

Careful consideration of the results shown in Fig. \ref{fig:all_pot_comp} makes it evident that the simulation results are not adequately described by the expressions for dust potential using the Yukawa potential or the Multipole Expansion potential given by equation \ref{eq:ivlevpot}.  The comparison between a 3D PIC simulation and the numerical solution of linearized electrostatic potential carried out in [\onlinecite{Ludwig2012}] showed that the charge and potential of the downstream grains can be significantly modified by wake effects in a system of several grains, and the wake pattern can show significant deviation from the potential determined from a linear combination of a wake behind a single grain in subsonic flows.  These results are supported by the results presented in Fig. \ref{fig:analytic_pot_comp_full_chain}, where it is seen that the linear superposition of single dust grain potentials deviates from the simulation of 20 dust grains for ion sound speeds as low as 0.07 M (corresponding to Case 4).  These findings also correlate with the comparison shown in Fig. \ref{fig:all_pot_comp}, where it is seen that the potential for the dust strings is not well represented by the models developed for isolated grains.

Many stable self-organized dust structures have been observed in ground-based and microgravity plasma environments, but the available models for electric potential have been found to be not well suited for describing the interactions between large numbers of dust grains.  Further description of the electric potential structure is needed, given that the dynamics of charged species within a complex plasma and the form of electric potential are inextricably linked, and will be the focus of a forthcoming study.  

\subsection{Dust Grain Charging}
\label{graincharge}
The time a dust grain takes to reach equilibrium charge ($\sim$60-100 $\mu s$) is longer than the time it takes for an ionization wave to pass ($\sim$50 $\mu s$ peak-to-peak), meaning that the actual dust charges will differ from those shown in Table \ref{tab:lengths}.  As a simple model for the dust grain charging and de-charging during the various phases in ionization waves, we may consider the dust grain to be a spherical capacitor and evaluate the charging and de-charging time as a dust grain transitions between the plasma conditions in Cases 2, 3, and 4.  Simulation results show that time to reach the new equilibrium charge after transitioning between cases was found to be 60-100 $\mu s$, with the longest time taken when transitioning from the conditions in Case 4 to the conditions in Case 2 (which has the longest ion plasma period, shown in Table \ref{tab:params}).

\begin{figure}[ht]
\includegraphics[width=\linewidth]{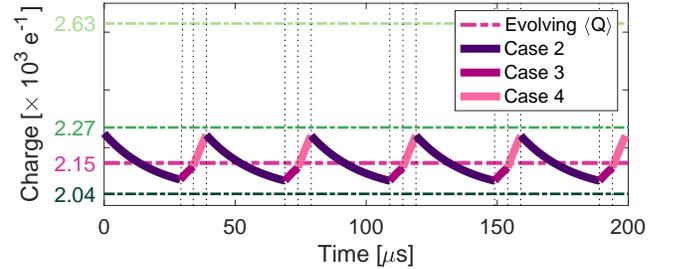}
\caption{Plot indicating the charge over time on a dust grain undergoing transitions between Cases 2, 3, and 4.  The color of the solid line segments in the charging curve indicates the particular transition, with dark purple corresponding to charging following the transition from Case 4 to Case 2, dark pink corresponding to charging following the transition from Case 2 to Case 3, and light pink corresponding to charging following the transition from Case 3 to Case 4.  The vertical dotted black lines indicate the transitions, and horizontal dot-dashed lines indicate the final equilibrium charge in dark green for Case 2 (2040 e$^-$), medium green for Case 3 (2270 e$^-$), light green for Case 4 (2630 e$^-$), and pink for the average charge resulting from transitions between Cases 2-4 (2150 e$^-$).}
\label{fig:dynamic_charging}
\end{figure}

As a rough approximation for the expected equilibrium charge for dust grains exposed to evolving plasma conditions due to ionization waves, a simple capacitor model was used to calculate the charge on a dust grain as a function of time as the grain charged or discharged in the plasma conditions associated with Cases 2-4.  This model was used to generate the piece-wise charging curve shown in Fig. \ref{fig:dynamic_charging}.  The final average charges in Cases 2-4 are shown as dot-dashed horizontal lines in shades of green, and the average equilibrium charge reached by the capacitor model in evolving conditions is shown as a dot-dashed horizontal pink line.  The piece-wise charging curve is shown by the thick solid line, with the color of the line segments corresponding to charging following the transition from Case 4 to Case 2 (dark purple), the transition from Case 2 to Case 3 (dark pink), and the transition from Case 3 to Case 4 (light pink).  Over 40 $\mu s$ (the approximate peak-to-peak time for the ionization waves), the charge fluctuates on the order of $\approx$ 200 e$^{-}$, with a time-averaged charge of 2150 e$^{-}$, which is lower than the charge predicted by Case 1.  This lower charge would result in smaller interparticle spacing, more closely matching that observed in the experimental data (Fig. \ref{fig:pk4_comp}).  However, it is important to note that this simple model is based upon constant conditions present within snapshots in time of dynamically varying plasma conditions, and as such may vary from the equilibrium charge found in an experimental environment.  In light of the impact fluctuating charges will have on the formation and stability of dust structure, future investigations utilizing time varying plasma parameters are needed to further quantify the impact ionization waves have on dust dynamics in the PK-4 environment.

\section{Conclusion}
\label{conclusion}
Varying conditions within the ionization waves present in the PIC-MCC simulation of conditions within the PK-4 ISS\cite{Hartmann2020} have a significant impact on the formation of dust chains.  The effect of fast-moving ionization waves on the formation of ordered dust particle chains and the resulting electric potential structure has been investigated by first identifying regions of interest within the time evolving plasma conditions (shown in Fig. \ref{fig:inputdatacomp}), labeled as the four cases: (1) \emph{Full Time Average}, where the values of each parameter are averaged over a time span (440 $\mu$s) encompassing several complete ionization waves; (2) \emph{Between $E_z$ Peaks}, where only the data in the interval between the large peaks in the axial electric field are considered, representing a region of minimum ion flow and a steady, constant plasma column; (3) \emph{Rising $E_z$ and Minimum $n_i$}, where the data within the rise of the axial electric field are considered, corresponding to an interval where electron and ion number densities are depleted; and (4) \emph{Full-Width Half-Maximum $E_z$ Peaks}, where the data is averaged over the FWHM of peaks in the axial electric field, representing the maximum ion flow.  The specific plasma parameters identified for each case (listed in Table \ref{tab:params}) were then modeled using DRIAD, a molecular dynamics simulation capable of resolving dust and ion motions on each of their respective timescales, and results for the ion density (Fig. \ref{fig:ionden}), electric potential (Fig. \ref{fig:totalpotential}), and analysis of the final dust structure (Fig. \ref{fig:gr_function}) have been presented for each of the four cases.

Although considered separately in this investigation, it is important to recall that the conditions of Cases 2, 3, and 4 represent different features within the ionization waves, with each case describing an interval of 10-20 $\mu s$ within a full ionization wave (with peak-to-peak interval of $\approx$ 40 $\mu$s), while the relevant dust dynamics occur on a timescale of $100$ $\mu$s.  As such, in the time-evolving experimental system, the decreased ion wake structure and diminished order for Case 2 (Fig. \ref{fig:ionden}b and \ref{fig:gr_function}b) is immediately followed by the enhanced wake and order found in Case 3 (Fig. \ref{fig:ionden}c and \ref{fig:gr_function}c).  As the ionization wave moves past a dust chain, conditions between ionization wave peaks (minimum ion velocities and maximum ion density) lead to an increase in ion current and subsequent decrease in the magnitude of charge on the dust grains, which allows dust grains to move closer together.  As the wave continues past the dust grains (with increased ion velocities and minimum ion density), the increased ion focusing leads to an enlarged region of positive potential between dust grains is the largest.  Even though the average charge on dust grains increases during this phase, the large positive potential in the strongly focused ion wake allows for smaller interparticle spacing relative to the shielding length, as well as increasing the grain alignment.  As shown in Fig.\ref{fig:gr_function} c-d, this leads to the largest order in the resulting dust chain.  The lower particle charge found using the capacitor charging model (Fig. \ref{fig:dynamic_charging}) allows for the smaller interparticle spacing, which more closely matches the experiment, while the enhanced ion wakes during the ionization waves contributes to the stability of the chain.

The balance between the various parameters is not captured by the full time average (Case 1), and as such, it is likely that the full time average does not provide an accurate representation of the conditions to be used to model dust dynamics.  Additionally, further work is needed in modeling the effect of ions through a modification of the potential of charged dust grains, as the current analytic models do not match the ion behavior.

\begin{acknowledgments}
Support for this work from NSF grant number 1740203, NASA grant number 1571701, and the U.S. Department of Energy, Office of Science, Office of Fusion Energy Sciences under award number DE-SC-0021334, and by the National Office for Research, Development, and Innovation of Hungary (NKFIH) via grant 134462 is gratefully acknowledged.  All authors gratefully acknowledge the joint ESA - Roscosmos "Experiment Plasmakristall-4" on-board the International Space Station. The microgravity research is funded by the space administration of the Deutsches Zentrum f\"{u}r Luft- und Raumfahrt eV with funds from the federal ministry for economy and technology according to a resolution of the Deutscher Bundestag under Grants No. 50WM1441 and No. 50WM2044.  A. M. Lipaev and A. D. Usachev were supported by the Russian Science Foundation Grant No. 20-12-00365 and participated in preparation of this experiment and its execution on board the ISS.  Color schemes used for figures 7, 11, and 13 were derived from Paul Tol's notes at (personal.sron.nl/$\sim$pault).

The authors gratefully acknowledge helpful ideas and suggestions made by the reviewers which led to interesting improvements to the manuscript.
\end{acknowledgments}

\section*{Data Availability Statement}
The data that support the findings of this study are available from the corresponding author upon reasonable request.

\bibliography{paper1_noMarkup}

\end{document}